\documentclass[aps,prl,twocolumn,superscriptaddress,preprintnumbers]{revtex4}

\usepackage{epsf,epsfig,amsmath,amssymb,amsfonts}
\usepackage{color}
\usepackage{slashed}

\newsavebox{\ns}
\newsavebox{\dbrane}

\def\be{\begin{equation}}
\def\ee{\end{equation}}
\def\bea{\begin{eqnarray}}
\def\eea{\end{eqnarray}}

\def\Dslash{\,\,{\raise.15ex\hbox{/}\mkern-12mu D}}
\def\Dbarslash{\,\,{\raise.15ex\hbox{/}\mkern-12mu {\bar D}}}
\def\delslash{\,\,{\raise.15ex\hbox{/}\mkern-9mu \partial}}
\def\delbarslash{\,\,{\raise.15ex\hbox{/}\mkern-9mu {\bar\partial}}}
\def\pslash{\,\,{\raise.15ex\hbox{/}\mkern-9mu p}}
\def\calDslash{\,\,{\raise.15ex\hbox{/}\mkern-12mu {\cal D}}}

\newcommand\diff{\mbox{d}}

\newcommand{\vol}{\mbox{vol}}

\newcommand{\nn}{\nonumber \\}

\newcommand{\dd}{\diff}

\DeclareMathOperator{\im}{Im}
\DeclareMathOperator{\re}{Re}

\allowdisplaybreaks

\begin{document}

\title{Large superconformal near-horizons from M-theory}

\preprint{DMUS--MP--16/01}         

\preprint{YITP-SB-2015-48}   

\preprint{APCTP Pre2016 - 008}

\vskip 1 cm

\author{\"O. Kelekci}
\affiliation{Department of Mathematics, Istanbul Technical University, Maslak 34469, Istanbul, Turkey}
 \affiliation{Department of Physics, Middle East Technical University, 06800, Ankara, Turkey}
 \author{Y. Lozano}
 \affiliation{Departamento de F\'isica, Universidad de Oviedo, Oviedo 33007, Spain}
 \author{J. Montero}
 \affiliation{Departamento de F\'isica, Universidad de Oviedo, Oviedo 33007, Spain}
 \author{E. \'O Colg\'ain}
 \affiliation{C. N. Yang Institute for Theoretical Physics, SUNY Stony Brook, NY 11794-3840, USA}
 \affiliation{Department of Mathematics, University of Surrey, Guildford GU2 7XH, UK}
 \affiliation{Asia Pacific Center for Theoretical Physics, Pohang 37673, Korea \\
 Department of Physics, Postech, Pohang 37673, Korea}
\author{M. Park}
\affiliation{School of Physics, Korea Institute for Advanced Study, Seoul 130-722, Korea}

\begin{abstract}
We report on a classification of supersymmetric solutions to 11D supergravity with $SO(2,2) \times SO(3)$ isometry, which are AdS/CFT dual to 2D CFTs with $\mathcal{N} = (0,4)$ supersymmetry. We recover the Maldacena, Strominger, Witten (MSW) near-horizon with small superconformal symmetry and identify a class of $AdS_3 \times S^2 \times S^2 \times CY_2$ geometries with emergent large superconformal symmetry. This exhausts known compact geometries. Compactification of M-theory on $CY_2$ results in a vacuum of 7D supergravity with large superconformal symmetry, providing a candidate near-horizon for an extremal black hole and a potential new setting to address microstates.
\end{abstract}

\maketitle

\setcounter{equation}{0}

\section{Introduction} \label{Introduction}
To obey the second law of thermodynamics, black holes must possess entropy, which Bekenstein \& Hawking showed is proportional to the area of the event horizon \cite{Bekenstein:1972tm}. This observation paved the way for the holographic principle and AdS/CFT \cite{Maldacena:1997re}. 
One of the earliest AdS/CFT calculations (it predates the conjecture) shows that asymptotic symmetries of gravity in $AdS_3$ correspond to the Virasoro algebra \cite{Brown:1986nw}, a feature of 2D CFTs. This observation together with the Cardy formula \cite{Cardy:1986ie} for the asymptotic growth of states for a CFT with central charge $c$ is enough to provide a microscopic derivation for the Bekenstein-Hawking (BH) entropy \cite{Strominger:1996sh, Strominger:1997eq}. For black holes with $AdS_3$ near-horizons, this methodology has been an incredible success, culminating in recent years in generalisations to extreme Kerr black holes \cite{Guica:2008mu}, potential astrophysical bodies \cite{extreme_BH}. 

However, Einstein's gravity is at best an effective description \cite{Donoghue:1994dn}, and the BH entropy is expected to be corrected in a candidate UV complete theory, such as M-theory. More concretely, compactifying the 6D M5-brane theory on a four-cycle in a Calabi-Yau three-manifold, $CY_3$, gives rise to the MSW CFT \cite{Maldacena:1997de}, with $\mathcal{N} = (0,4)$ supersymmetry at low energies. The corresponding black hole exhibits the near-horizon $AdS_3 \times S^2 \times CY_3$, and sub-leading corrections to the BH entropy have been shown to perfectly match corrections to the central charge \cite{Maldacena:1997de, Freed:1998tg}. 

The MSW CFT exhibits \textit{small} superconformal symmetry \cite{Ademollo:1975an} with an $SU(2)$ R symmetry that is manifest in the two-sphere in the dual geometry. Since other superconformal symmetries exist \cite{Schoutens:1988ig, Sevrin:1988ew, Ali:1999ut}, a rich class of AdS/CFT geometries can be expected, e. g.  \cite{D'Hoker:2008ix}. In this letter, we identify a new class of M-theory vacua $AdS_3 \times S^2 \times S^2 \times CY_2$, implying the existence of a distinct class of 2D $\mathcal{N} = (0,4)$ CFTs with \textit{large} superconformal symmetry and R symmetry $SU(2) \times SU(2)$. We recall that CFTs with large superconformal symmetry remain largely enigmatic. While constructions based on string theory, such as $AdS_3 \times S^3 \times S^3 \times S^1$ \cite{Boonstra:1998yu, deBoer:1999gea}, exist, contrary to small superconformal CFTs, interpretation as a symmetric product CFT is problematic \cite{Gukov:2004ym}. This issue continues to attract exciting new holographic proposals  \cite{Gaberdiel:2013vva, Tong:2014yna}, against a backdrop where we have witnessed a deeper understanding of the r\^{o}le of integrability \cite{Babichenko:2009dk, OhlssonSax:2011ms, Sax:2012jv}. 

More concretely, we report the results of a complete classification of supersymmetric solutions to 11D supergravity, the low-energy description of M-theory, where we assume $SO(2,2) \times SO(3)$ isometry, i. e. warped $AdS_3 \times S^2 \times \mathcal{M}_6$ spacetime, with $\mathcal{M}_6$ being an $SU(2)$-structure manifold. Since 2D $\mathcal{N} = (0,4)$ SCFTs are expected to exhibit at least an $SU(2)$ isometry, corresponding to the R symmetry, this is a minimal requirement. One may contemplate a distinct class where the $SU(2)$ R symmetry is realised as a squashed three-sphere, $\tilde{S}^3$, but such an ansatz would preclude the MSW geometry. Indeed, non-compact $AdS_3 \times \tilde{S}^3 \times S^2 \times T^3$ geometries, generated via non-Abelian T-duality \cite{Sfetsos:2010uq, Lozano:2011kb}, were identified recently in ref. \cite{Lozano:2015bra}. It is also an immediate corollary of this work that compact $AdS_3 \times \tilde{S}^3 \times \tilde{S}^3 \times T^2$ geometries with $\mathcal{N} = (0,4)$ supersymmetry may be generated through TsT transformations \cite{Lunin:2005jy}. 

Our results pertain to general warped $AdS_3 \times S^2$ spacetimes and are not intended to apply to all M-theory geometries dual to 2D $\mathcal{N} = (0,4)$ SCFTs. Within our assumptions, we prove that $\mathcal{M}_6$ is either $CY_3$, thus recovering the MSW geometry, or it possesses an additional $SU(2)$ R symmetry that emerges from the supersymmetry analysis. Truncating the emergent $SU(2)$ to $U(1)$, we recover a known class \cite{Gauntlett:2006ux, Kim:2007hv} of spacetimes with $SU(2) \times U(1)$ isometry \footnote{Indeed, the existence of the $SU(2) \times U(1)$ isometry is a puzzle, since typically emergent isometries are R symmetries, but $SU(2) \times U(1)$ does not fit with a known $\mathcal{N}=4$ superconformal algebra. Recently, the first explicit example of a geometry in this class was identified \cite{Lozano:2015bra} and the emergent $U(1)$ symmetry shown not to be an R symmetry, but rather the M-theory circle.}. 

The existence of a class of $AdS_3 \times S^2 \times S^2 \times CY_2$ solutions to 11D supergravity, with 2D $\mathcal{N} = (0,4)$ SCFT duals, comes somewhat as a surprise. In the case where $CY_2 = T^4$, it was shown long ago that there are geometries related through T-duality to well-known $AdS_3 \times S^3 \times S^3 \times S^1$ solutions in 10D \cite{Boonstra:1998yu}. When $CY_2 = K_3$, the class appears new. It did not feature in a study of wrapped M5-brane geometries \cite{Gauntlett:2006ux}. More recently, M-theory geometries dual to 2D $\mathcal{N} = (0,2)$ SCFTs have been discussed, but where supersymmetry is enhanced to $\mathcal{N} = (0,4)$, the geometry is either MSW \cite{Gauntlett:2006qw,Bah:2015nva}, or no good $AdS_3$ vacuum exists \cite{Benini:2013cda, Karndumri:2015sia}. Moreover, it is expected that M-theory on $K_3$ is dual to heterotic string theory on $T^3$ \cite{Hull:1994ys}, a statement that can be made precise in the supergravity limit \cite{Lu:1998xt}. Despite this, in a recent classification of heterotic supergravity \cite{Beck:2015gqa}, the only compact, regular solutions with eight supersymmetries are shown to be $AdS_3 \times S^3 \times CY_2$ \footnote{It has been suggested that the dilaton is singular in the near-horizon of intersecting NS5-branes and this violates the partial integration argument of ref. \cite{Beck:2015gqa}. We thank G. Papadopoulos for correspondence.}. 

It can be expected our simply-stated results will be of interest to anyone studying the holography of 2D $\mathcal{N} = (0,4)$ CFTs.

\vspace{-2mm}
\section{$SO(2,2) \times SO(3)$-invariant spacetimes}
\vspace{-2mm}

We recall that bosonic sector of 11D supergravity consists of a metric, $g$, and a three-form potential, $C$, with four-form field strength, $G = \dd C$. The equations of motion follow from the action 
\be
\label{action}
S = \frac{1}{2 \kappa^2} \int * R - \frac{1}{2} G \wedge * G - \frac{1}{6} C \wedge G \wedge G. 
\ee
Supersymmetric solutions satisfy the Killing spinor equation (KSE): 
\be
\label{KSE}
\nabla_{M} \eta + \frac{1}{2 88} \left[ \Gamma_{M}^{~NPQR} - 8 \delta_{M}^{~N} \Gamma^{PQR} \right] G_{NPQR} \eta = 0,  
\ee
where $M, N = 0, \dots, 10$, $\nabla_{M} \eta \equiv \partial_{M} \eta + \frac{1}{4} \omega_{M NP} \Gamma^{NP} \eta$, with spin connection $\omega$, and $\eta$ is a Majorana Killing spinor. It is well-known that the Einstein equation is implied by the KSE once the Bianchi identity, $\dd G = 0$, and equation of motion for $C$ hold \cite{Gauntlett:2002fz}. 

2D $\mathcal{N}  = (0,4)$ CFTs enjoy both $SO(2,2)$ conformal symmetry and $SU(2) \simeq SO(3)$ R symmetry, which motivates the general ansatz
\bea
\label{11D_ansatz}
\dd s^2  &=& e^{2A} \left[ \frac{1}{m^2} \dd s^2(AdS_3) + e^{2 B} \dd s^2(S^2) + \dd s^2(\mathcal{M}_6) \right], \nn
G  &=& \frac{1}{m^3} \vol (AdS_3) \wedge \mathcal{A} + \vol(S^2) \wedge \mathcal{H} + \mathcal{G}, 
\eea
where $m$ is the inverse $AdS_3$ radius, $A, B$ denote scalar warp factors and $\mathcal{A}, \mathcal{H}, \mathcal{G}$ are respectively closed one, two and four-forms. The curvatures of symmetric spaces are canonically normalised and fields depend only on the coordinates of the internal 6D Riemannian manifold $\mathcal{M}_6$. 

In order to characterise the internal space and the fields, we decompose the 11D gamma matrices \cite{Colgain:2010wb}
\bea
\Gamma_{\mu} &=& \tau_{\mu} \otimes \sigma_3 \otimes \gamma_7, ~~
\Gamma_{\alpha} = 1_2 \otimes \sigma_{\alpha} \otimes \gamma_7, \nn
\Gamma_{m} &=& 1_2 \otimes 1_2 \otimes \gamma_m,   
\eea
and 11D Killing spinor, 
\be
\label{11D_spinor_decomp}
\eta = \psi \otimes e^{A/2} \left[ \chi_+ \otimes \epsilon_+ + \chi_- \otimes \epsilon_- \right], 
\ee
where $\mu = 0, 1, 2$ label $AdS_3$ directions, $\alpha = 1, 2$ denote those of $S^2$, $m=1, \dots, 6$ correspond to $\mathcal{M}_6$ and we define $\gamma_7 \equiv i \gamma_{123456}$. $\psi$ is a solution to the $AdS_3$ KSE, $\nabla_{\mu} \psi = \frac{1}{2} \tau_{\mu} \psi$, resulting in Poincar\'e spinors of definite chirality, while $\chi_{\pm}$ denote an $SU(2)$-doublet satisfying the KSE on $S^2$, $\nabla_{\alpha} \chi_{\pm} = \pm \frac{i}{2} \sigma_{\alpha} \chi_{\pm}$, with $\chi_- = \sigma_3 \chi_{+}$. It is a common feature of Refs. \cite{Kim:2007hv, Colgain:2010wb, Lin:2004nb} that the Majorana condition is not manifest, however conjugate spinors, $\eta^{c}$, may easily be constructed e. g. \cite{Chen:2010jga}. Following the decomposition through, one determines the effective 6D KSE equations in terms of $\epsilon_{\pm}$ \cite{Colgain:2010wb} and recasts them in terms of conditions on differential forms \cite{Tod:1983pm}, which we illustrate later. 

We stress that there is $\textit{a priori}$ no relation between $\epsilon_{\pm}$, even if one is to be expected \footnote{Naive supersymmetry counting works here: four $AdS_3$ spinors and two $S^2$ spinors already give eight supersymmetries, so we just expect one from $\mathcal{M}_6$.}. In related work, Ref. \cite{Lin:2004nb} simplified the problem by omitting a term in the four-form flux, which enabled a simplification of the KSE analysis, before showing that the omitted term could not be reconciled perturbatively. This term was later ruled out in general \cite{OColgain:2010ev}. In the current setting, this simplification involves fixing $\mathcal{A} = \mathcal{G} =0$. However, since geometries with non-zero $\mathcal{A}, \mathcal{H}, \mathcal{G}$ can be generated via T-duality \cite{Lozano:2015bra}, this simplification is difficult to motivate. 

\vspace{-2mm}
\section{Supersymmetry conditions} 
\vspace{-2mm}
We review the salient conditions on bilinears, defined in the appendix, which we construct from spinors $\epsilon_{\pm}$ \cite{Colgain:2010wb}, which encapsulate the local supersymmetry conditions we must solve. Firstly, supersymmetry demands that the following bilinears vanish \cite{Colgain:2010wb}, 
\be
\label{trivial_scalars}
W^- = X^+ = \re(Y) = \tilde{Z} = 0. 
\ee
Moreover, the remaining bilinears are constrained
\bea
\label{scalar_constraint1}
2 m V^+ &+& e^{-B} \im(Y) \nn 
&=& \frac{e^{-3A}}{2} \left[ \frac{1}{2!} \im(L^3)_{mn} (*_6 \mathcal{G})^{mn}+K^{+}_m \mathcal{A}^m \right], \\
\label{scalar_constraint2} \tilde{Y} &=& - \frac{i}{2 m e^{B}} W^+, \quad Z = - \frac{i}{2 m e^{B}} X^-. 
\eea
Thus, there are only three real scalars, $V^{\pm}$, $W^+$, and one complex scalar, $X^-$, which can be independent. 

From the vector spinor bilinears, one can identify four \textit{real} Killing vectors on $\mathcal{M}_6$ \cite{Colgain:2010wb}, three of which, $\im(\tilde{K}^3), \re(K^4)$ and $\im(K^4)$ extend to symmetries of the overall solution (\ref{11D_ansatz}). In contrast, the $S^2$ warp factor (also $\mathcal{H}$) depends on $\tilde{K}^+$, thus hinting at spacetimes with larger superisometry groups \footnote{We have checked that $\tilde{K}^+$ is non-zero for maximally supersymmetric $AdS_7 \times S^4$.}. Thankfully, $\tilde{K}^+$ may be truncated out consistently provided $V^{-} = 0$, i. e. for 6D spinors $\epsilon_{\pm}$ with equal norm. Henceforth, we consider $\epsilon^{\dagger}_{+} \epsilon_+ = \epsilon_-^{\dagger} \epsilon_-$, so that $V^- = \tilde{K}^+ = 0$. 

The scalars satisfy differential constraints \cite{Colgain:2010wb},
\bea
\label{diff1} \dd  V^+ &=& 0, \\
\label{diff2} \dd [e^{-B} \im(Y) ] &=& 0, \\
\label{diff3} e^{-3A} \dd [ e^{3A} X^- ] &=& - 2 m \tilde{K}^4, \\
\label{diff4} e^{-3A} \dd [e^{3A} W^+ ] &=& 2 m \re (K^3), 
\eea
while the vectors must satisfy 
\bea
\label{diff6} \dd [ e^{3 A + B} K^- ] &=& - e^{-B} \im(Y) \mathcal{H} + e^{3A} \tilde{L}^1, \\
\label{diff8} \dd[ e^{6A+B} \re(\tilde{K}^3)] &=& - e^{3A+B} \im(Y) *_6 \mathcal{G}  \nn
&-& e^{3A+B} \mathcal{A} \wedge K^- - e^{6A} \im(L^3), \\
\label{diff9} \dd [e^{6A+2 B} \im(\tilde{K}^3)] &=& - e^{3A} W^+ \mathcal{H} + 2 m e^{6A+2 B} L^1 \nn &+& 2 e^{6A+B} \re(L^3), \\
\label{diff10} \dd [ e^{6A+2B} K^4] &=& i e^{3A} X^- \mathcal{H} + 2 m e^{6A+2B} L^6 \nn &-& 2 i e^{6A +B} \tilde{L}^4. 
\eea
We have removed all trivial bilinears and conditions that play no r\^{o}le in our analysis \footnote{We have corrected a sign typo in (\ref{diff8}).}. With $\tilde{Y}$ pure imaginary from (\ref{scalar_constraint2}), and $\tilde{K}^+$ zero, $\mathcal{A}$ and $\mathcal{G}$ are fully determined in terms of bilinears: 
\bea
\label{AG} \mathcal{A} &=& \frac{2m e^{3 A}}{V^+} K^+, \quad  \mathcal{G} = \frac{2m e^{ 3A}}{V^+} *_6 \im(L^3). 
\eea
This ends our review of the supersymmetry conditions of Ref. \cite{Colgain:2010wb}. We will now solve the conditions by evoking $G$-structures to characterise the internal manifold $\mathcal{M}_6$. 

We introduce two unit-norm, chiral spinors, $\xi_i$, which are orthogonal, $\xi_{i}^{\dagger} \xi_j = \delta_{ij}$. Each chiral spinor individually defines an $SU(3)$-structure. To see this, we introduce projection conditions
\bea
\label{proj_x1}
\gamma_{12} \xi_1 = \gamma_{34} \xi_1 = \gamma_{56} \xi_1 &=& i \xi_1,  ~~\Rightarrow~~ \gamma_7 \xi_1 = \xi_1, \nn
-\gamma_{135} \xi_1 &=& \xi_1^*, 
\eea
permitting us to specify the $SU(3)$-structure through a two-form $J^{(3)}_i = - \frac{i}{2} \xi_i^{\dagger} \gamma_{mn} \xi_i \, e^{m} \wedge e^n$ and $(3,0)$-form $\Omega^{(3)}_i = - \frac{1}{3!} \xi^{T}_i \gamma_{mnp} \xi_i \, e^m \wedge e^n \wedge e^p$. With the second spinor, $\gamma_5 \xi_2^{*} = \xi_1$, whose projection conditions follow from (\ref{proj_x1}), we can define two canonical $SU(3)$-structures, with forms $(J^{(3)}_1, \Omega^{(3)}_1)$ and ($J^{(3)}_2, \Omega^{(3)}_2$), or equivalently, a canonical $SU(2)$-structure, which is specified by three two-forms and two one-forms: 
\bea
\label{harmonic} J^{\alpha} &=& - \frac{i}{4} (\sigma^{\alpha})^{ij} \xi_{i}^{\dagger} \gamma_{mn} \xi_j \, e^{m} \wedge e^n, \\
 K^1 + i K^2 &=& -\frac{1}{2} \epsilon^{i j} \xi^{T}_i \gamma_m \xi_j \, e^m, 
\eea
where $\sigma^{\alpha}, \alpha =1, 2, 3$ denote Pauli matrices. In general, we expand $\epsilon_{\pm}$ in terms of the chiral spinors and their conjugates
\bea
\label{ep_em}
\epsilon_+ &=& \alpha_1 \xi_1 + \alpha_2 \xi_1^* + \alpha_3 \xi_2 + \alpha_4 \xi_2^*, \nn
\epsilon_- &=& \beta_1 \xi_1 + \beta_2 \xi_1^* + \beta_3 \xi_2 + \beta_4 \xi_2^*,
\eea
where $\alpha_i, \beta_i \in \mathbb{C}$. Modulo phases of $\xi_i$, these are the most general spinors consistent with $SU(2)$-structure. 

\vspace{-2mm}
\section{$SU(2)$-structure manifolds}
\vspace{-2mm}

As a warm-up, we consider $SU(3)$-structure manifolds by simply eliminating, $\alpha_3, \alpha_4, \beta_3, \beta_4$, so that only $\xi_1$ remains in (\ref{ep_em}). We recall that $SU(3)$-structure manifolds are classified according to five torsion classes $W_i$, \cite{Chiossi:2002tw}. 
We will now demonstrate that all torsion classes vanish, so Calabi-Yau is the only $\mathcal{M}_6$ with $SU(3)$-structure. 

One can use the constraints from vanishing scalars (\ref{trivial_scalars}) to infer, $\epsilon_- = \pm i \epsilon_+$, where $\epsilon_+$ need not be chiral. If it is chiral, the argument reverts to Ref. \cite{Colgain:2010wb}; if non-chiral, since $K^+ = \im(L^3)= 0$, $\mathcal{A}$ and $\mathcal{G}$ also are zero. Next, from (\ref{scalar_constraint1}),  we deduce $ 2m e^{B} = \mp 1$, and from (\ref{diff6}), $\mathcal{H} = \pm e^{3A}/(2 m) J^{(3)}~(V^+=1)$. Further differentiating (\ref{diff6})-(\ref{diff10}), we can directly confirm that $\dd J^{(3)} = \dd \Omega^{(3)} = 0$. 

We now turn to the generic case. Evaluating the vector bilinears in terms of $\alpha_i, \beta_i$ using (\ref{ep_em}), we find that $K^-$ and $\tilde{K}^-$ are orthogonal allowing us without loss of generality to align them with the $e^5$, $e^6$ axes of the internal space. Moreover, imposing (\ref{trivial_scalars}), and $V^- = \tilde{K}^+ = 0$, we determine the following relations: 
\bea
\beta_1 &=& - \frac{\left[ \alpha_3 (\alpha_4^2 + \beta_4^2) + \alpha_1 (\beta_2 \beta_4 + \alpha_2 \alpha_4)\right]}{(\beta_2 \alpha_4 - \alpha_2 \beta_4)} , \nn
\beta_3 &=&  \frac{ \left[ \alpha_1 (\alpha_2^2 + \beta_2^2) + \alpha_3 (\beta_2 \beta_4 + \alpha_2 \alpha_4)\right]}{(\beta_2 \alpha_4 - \alpha_2 \beta_4)},  \nn
 i f_1 &=& \beta_2 \alpha_2^* + \beta_4 \alpha_4^* , \quad f_2 = \alpha_1^* \alpha_4 + \alpha_2^* \alpha_3, 
\eea
where $f_i \in \mathbb{R}$ are yet to be determined. With $SU(2)$-structure, it follows from $\tilde{K}^+ = 0$ that $K^+ = 0$ and, as a result of (\ref{AG}), $\mathcal{A} = 0$, i. e. no electric flux. As another consequence of these relations, we discover $\re(\tilde{K}^3 ) = 0$, which through (\ref{diff8}) and (\ref{AG}) leads to the constraint 
\be
Y = - \frac{i}{2 m e^{B}} V^+. 
\ee
Since $V^+$ is a constant, so too is $e^{B}$ through (\ref{diff2}). 

We can now combine this with (\ref{scalar_constraint2}) to find that 
\be
\label{f2}
\left[ \beta_4 + \frac{i}{2 m e^{B}} \alpha_4\right] (f_2 - \alpha_2^* \alpha_3 - \alpha_3^* \alpha_2) = 0. 
\ee
If we impose the vanishing of the first bracket, through the constraints it follows that $\mathcal{G} = 0$ and $\beta_i = - i \alpha_i$, i. e. $\epsilon_- = - i \epsilon_+$, so that we recover Calabi-Yau. To find something new, we impose the second condition, which implies $K^- = \tilde{K}^- = 0$. We recall that these are the original vectors that we aligned with the axes, so now we have the freedom to choose $K^3$ and $\tilde{K}^3$, which are orthogonal, and rotate them to align with the axes. Doing so, we find it is possible to solve for all the spinor coefficients so that our constraints are satisfied: 
\bea
\alpha_1 &=& \sqrt{V^+} \cos \frac{\zeta}{2} \cos \frac{\theta}{2} e^{i \varphi_1}, \quad \alpha_2 = \sqrt{V^+} \sin \frac{\zeta}{2} \cos \frac{\theta}{2} e^{i \varphi_2}, \nn
\alpha_3 &=& \sqrt{V^+} \cos \frac{\zeta}{2} \sin \frac{\theta}{2} e^{i \varphi_3}, \quad \alpha_4 = \sqrt{V^+} \sin \frac{\zeta}{2} \sin \frac{\theta}{2} e^{i \varphi_4}, \nn
\beta_1 &=& \frac{1}{2} \left( \frac{L}{R_2} \cot \frac{\zeta}{2} \alpha_4 - i \frac{L}{R_1} \alpha_1 \right), \quad \beta_4 = \frac{\beta_2}{\beta_1^*} \beta_3^*, \nn
\beta_3 &=&   -\frac{1}{2} \left( \frac{L}{R_2} \cot \frac{\zeta}{2} \alpha_2 + i \frac{L}{R_1} \alpha_3 \right), \quad \beta_2 = - \frac{\alpha_2}{\alpha_1^*} \beta_1^*, 
\eea
where $\varphi_1 + \varphi_2 = \varphi_3 + \varphi_4$ and we have redefined  $m = L^{-1}, R_1 = e^{B}, R_2 = e^{B}/\sqrt{ 4 m^2 e^{2 B}-1}$. With these expressions, we determine $W^+ = V^+ \cos \zeta, X^- = V^+ \sin \zeta e^{i \varphi_1 + i \varphi_2}$ and solve (\ref{diff3}) and (\ref{diff4}) to show the warp factor $e^{A}$ is a constant and 
\be
e^{5} = - R_2 \dd \zeta, \quad e^6 = - R_2 \sin \zeta \dd \chi,  
\ee
where we have defined $\dd \chi = \dd (\varphi_1 + \varphi_2)$. This allows us to identify the one-forms dual to the Killing vectors,
\bea
\im(\tilde{K}^3) &=& -\frac{L V^+}{2} \sin^2 \zeta \dd \chi, \nn
K^4 &=& - \frac{L V^+}{2} e^{i \chi} \left( \dd \zeta + i \cos \zeta \dd \chi \right), 
\eea
which correspond to an emergent $SU(2)$. We can ensure the Killing vectors are canonically normalised through the choice  $V^+ = 2 R_2^2 /L$. Solving the remaining supersymmetry conditions, one arrives at the conclusion that $\chi$ aside, the other angular parameters are constant, with $\mathcal{M}_6$ being a direct-product of  $S^2$ and $CY_2$, more concretely $T^4$ or $K_3$. The final expression for the four-form flux reads
\be
G = \frac{2 e^{3A}}{L V^+} \left[ - R_1^2 \tilde{L}^1 \wedge \vol(S^2) +  *_6 \im(L^3) \right]. 
\ee
It is easy to check that the equations of motion are satisfied, in line with expectations \cite{Gauntlett:2002fz}. We also see that both $\xi_1, \xi_2$ and conjugates need to appear in the spinor. This may be contrasted with the spinor considered in Ref. \cite{Gauntlett:2004zh}, which is not the most general, and would appear to preclude this outcome.  For this reason, setting $\beta_4 = \alpha_4 = 0$ in (\ref{f2}), one recovers the results of existing classifications \cite{Gauntlett:2006ux, Kim:2007hv}. Setting $A=0$, since the overall warp-factor is constant,  we can confirm the radii satisfy 
\be
\frac{4}{L^2}  = \frac{1}{R_1^2} + \frac{1}{R_2^2}.  
\ee
The ratio between $S^2$ radii, $\alpha$, corresponds to the supergroup $D(2, 1;\alpha)$, with bosonic subgroup $SL(2, \mathbb{R}) \times SU(2) \times SU(2)$.   

To establish the connection to minimal ungauged supergravity in 7D \cite{Townsend:1983kk}, we exploit the following consistent Kaluza-Klein reduction ansatz: 
\bea
\dd s^2_{11} &=& e^{- \frac{8}{5} B} \dd s^2_7 + e^{2 B} \dd s^2 (CY_2), \nn
G &=& F  + \sum_{a=1}^3 F^a \wedge J^a, 
\eea
where $J^a$ denote the three self-dual harmonic two-forms of $CY_2$, $B$ is a scalar and $F$ and $F^a$ are respectively field strengths corresponding to a three-form and one-form potentials, $F = \dd C, F^a = \dd A^a$. The resulting action in Einstein frame in 7D is 
\bea
\mathcal{L}_7 &=& R \vol_7 - \frac{36}{5} \dd B \wedge *_7 \dd B - \frac{1}{2} e^{ \frac{24}{5} B} F \wedge *_7 F \nn 
&-& e^{-\frac{12}{5} B} F^a \wedge *_7 F^a - F \wedge F^a \wedge A^a. 
\eea
To cast the action in the original notation of ref. \cite{Townsend:1983kk}, one should employ the following redefinitions: 
\bea
B = \frac{\sqrt{5}}{6} \phi, \quad F_{\textrm{us}} = \sqrt{2} F_{\textrm{them}}, \quad F_{\textrm{us}}^a = \sqrt{2} F^{a}_{\textrm{them}}. 
\eea

\vspace{-2mm}
\section{Discussion}
\vspace{-2mm}
We have initiated a classification of all solutions to 11D supergravity with $SO(2,2) \times SO(3)$ isometry. This is the simplest geometric signature of a supergravity solution dual to a 2D CFT with $\mathcal{N} = (0,4)$ supersymmetry, including the MSW CFT. In the process, we have identified a novel class of near-horizon geometries in M-theory with large superconformal symmetry. Compactifying M-theory on $CY_2$, we identify a resulting $AdS_3 \times S^2 \times S^2$ vacuum to 7D supergravity, thus providing a candidate near-horizon for an extremal black hole and a potential new controlled setting to count black hole microstates.  

The M-theory geometry provides a unifying description of well-known $AdS_3 \times S^3 \times S^3 \times S^1$ geometries of Type II string theory through T-duality \cite{Lozano:2015bra} and heterotic vacua via M-theory/heterotic duality \cite{Hull:1994ys}. A careful treatment of the central charge reveals the expected form of a large superconformal algebra \cite{inprogress}
\be
c\sim \frac{k^+ k^-}{k^++k^-},
\ee
with affine $SU(2)^\pm$ current algebras at levels $k^\pm$ related to the quantised charges, yet where $c \sim N^2$, for large charge $N$, and not the more usual $c \sim N^3$ of geometries corresponding to M5-branes. 

Our work has two interesting implications. Firstly, it is striking that the $AdS_3 \times S^2 \times S^2 \times CY_2$ geometries are not identifiable as $AdS_3$ limits of wrapped M5-branes \cite{Gauntlett:2006ux}. This suggests the M5-brane picture is novel and motivates further study to understand anomaly in-flow \cite{Freed:1998tg}. Secondly, as we have shown, since 11D supergravity compactifies on $CY_2$ to 7D minimal supergravity, the $AdS_3 \times S^2 \times S^2$ solution hints at being the near-horizon of an extremal black hole. While such solutions have in principle been classified \cite{Cariglia:2004qi}, we are not aware of a near-horizon uniqueness theorem in 7D, cf. \cite{Emparan:2008eg}. Assuming a black hole exists, strong parallels to the MSW case, with M-theory compactified on Calabi-Yau, are expected to facilitate a microscopic derivation of the entropy. Since the small superconformal algebra is recovered from the large one through a decompactification of a two-sphere, it is tempting to speculate that contact with the MSW results may be made in the same limit. 

Lastly, we remark that we have assumed $SU(2)$-structure, and more general solutions with identity structure are known to exist \cite{Lozano:2015bra}.  We hope to extend the classification to consider more general internal manifolds in future work \cite{inprogress}. 

\section*{Acknowledgements}	
We have enjoyed discussions with D. Chow, D. Giataganas, K. M. Lee, S. Nam, G. Papadopoulos, C. Park, M. Ro\v{c}ek, D. C. Thompson, H. Yavartanoo \& P. Yi. O.K is partially supported by TUBITAK Grant No:114F321. Y.L. and J.M. are partially supported by the Spanish Ministry of Economy and Competitiveness grant FPA2012-35043-C02-02 and the EU-COST Action MP1210. J.M. is supported by the FPI fellowship BES-2013-064815 linked to the previous project, and is grateful for the warm hospitality extended by YITP, SUNY Stony Brook, where part of this work was done.
E \'O C is supported by the Marie Curie award PIOF-2012-328625 T-DUALITIES and wishes to thank APCTP and KIAS for hospitality.  M. Park is supported by TJ Park Science Fellowship of POSCO TJ Park Foundation. 

\appendix 

\section{Spinor bilinears}
In our conventions, the 6D gamma matrices are Hermitian $\gamma^{\dagger}_{m} = \gamma_m$ and anti-symmetric $\gamma^{T}_m = - \gamma_m$. Consistent with the symmetries of the gamma matrices, given $\epsilon_{\pm}$, we can define an exhaustive set of scalar 
\bea
\label{scalar_bilinears}
V^{\pm} &=& \frac{1}{2} ( \epsilon_+^{\dagger} \epsilon_+ \pm \epsilon_-^{\dagger} \epsilon_-), \nn
W^{\pm} &=& \frac{1}{2} ( \epsilon_+^{\dagger} \gamma_7 \epsilon_+ \pm \epsilon_-^{\dagger} \gamma_7 \epsilon_-), \nn
X^{\pm} &=& \frac{1}{2} ( \epsilon_+^{T} \epsilon_+ \pm  \epsilon_-^{T} \epsilon_-), \nn
Y &=& \epsilon_+^{\dagger} \epsilon_-, \quad \tilde{Y} = \epsilon_+^{\dagger} \gamma_7 \epsilon_-, \nn
Z &=& \epsilon_+^{T} \epsilon_-, \quad \tilde{Z} = \epsilon_+^{T} \gamma_7 \epsilon_-, 
\eea
and vector spinor bilinears: 
\bea
K^{\pm}_{m} &=& \frac{1}{2} (\epsilon_+^{\dagger} \gamma_m \epsilon_+ \pm \epsilon_-^{\dagger} \gamma_m \epsilon_-), \nn
\tilde{K}^{\pm}_{m} &=& \frac{i}{2}( \epsilon_+^{\dagger} \gamma_m \gamma_7 \epsilon_+ \pm \epsilon_-^{\dagger} \gamma_m \gamma_7 \epsilon_- ), \nn
K^3_{m} &=& \epsilon_+^{\dagger} \gamma_m \epsilon_-, \quad \tilde{K}^3_{m} = \epsilon_{+}^{\dagger} \gamma_m \gamma_7 \epsilon_-, \nn 
K^4_{m} &=& \epsilon_+^{T} \gamma_{m} \epsilon_-, \quad \tilde{K}^4_m = \epsilon_+^{T} \gamma_m \gamma_7 \epsilon_-, 
\eea
where factors of $i$ ensure vectors are real. We define the following two-forms:
\bea
L^1_{mn} &=& \frac{i}{2} (\epsilon_+^{\dagger} \gamma_{mn} \epsilon_+ + \epsilon_-^{\dagger} \gamma_{mn} \epsilon_-), \nn 
\tilde{L}^1_{mn} &=& \frac{i}{2}( \epsilon_+^{\dagger} \gamma_{mn} \gamma_7 \epsilon_+ + \epsilon_-^{\dagger} \gamma_{mn} \gamma_7 \epsilon_- ), \nn
L^3_{mn} &=& \epsilon_+^{\dagger} \gamma_{mn} \epsilon_-, \quad \tilde{L}^4_{mn} = \epsilon_+^{T} \gamma_{mn} \gamma_7 \epsilon_-, \nn
L^6_{mn} &=& \frac{1}{2}( \epsilon_+^{T} \gamma_{mn} \gamma_7 \epsilon_+ - \epsilon_-^{T} \gamma_{mn} \gamma_7 \epsilon_- ),
\eea
where notation follows Ref. \cite{Colgain:2010wb}.

\end{document}